\documentclass[cernpreprint, atlasdraft=false, texlive=2023, USenglish, orcidlogo, texmf, orcidlogo]{atlasdoc}
 
\usepackage{atlaspackage}
\usepackage[block=space,backref=false]{atlasbiblatex}
 
\usepackage[hion=true]{atlasphysics}

\graphicspath{{logos/}{figures/}}
 
\usepackage{ANA-STDM-2019-19-PAPER-defs}

\AtlasTitle{Observation of the $\gamma\gamma\to\tau\tau$ process in Pb+Pb collisions and constraints on the $\tau$-lepton anomalous magnetic moment with the ATLAS detector}
 
\AtlasVersion{3.0}
 
\AtlasAbstract{
This Letter reports the observation of $\tau$-lepton pair production in ultraperipheral lead--lead collisions, $\text{Pb+Pb} \to \text{Pb}(\gamma\gamma \to \tau\tau)\text{Pb}$, and constraints on the $\tau$-lepton anomalous magnetic moment, $a_\tau$.
The dataset corresponds to an integrated luminosity of 1.44 nb$^{-1}$ of LHC Pb+Pb collisions at $\sqrt{s_{_\text{NN}}} = 5.02$~TeV recorded by the ATLAS experiment in 2018.
Selected events contain one muon from a $\tau$-lepton decay, an electron or charged-particle track(s) from the other $\tau$-lepton decay, little additional central-detector activity, and no forward neutrons.
The $\gamma\gamma \to \tau\tau$ process is observed in Pb+Pb collisions with a significance exceeding 5~standard deviations, and a signal strength of $\mu_{\tau\tau} = 1.03^{+0.06}_{-0.05}$ assuming the Standard Model value for $a_\tau$.
To measure $a_\tau$, a template fit to the muon transverse-momentum distribution from $\tau$-lepton candidates is performed, using a dimuon ($\gamma\gamma \to \mu\mu$) control sample to constrain systematic uncertainties.
The observed 95\% confidence-level interval for $a_\tau$ is $-0.057<a_\tau <0.024$.
}

\AtlasRefCode{STDM-2019-19}
 
\PreprintIdNumber{CERN-EP-2022-079}

\AtlasJournalRef{Phys.\ Rev.\ Lett. 131 (2023) 151802}
\AtlasDOI{10.1103/PhysRevLett.131.151802}

\hypersetup{pdftitle={ATLAS document},pdfauthor={The ATLAS Collaboration}}
 
\begin{document}
 
\maketitle

 
Measurements of the anomalous magnetic moment, $a_\ell = \frac{1}{2}(g_\ell-2)$, of charged leptons $\ell$ (electrons, muons, and $\tau$-leptons) are cornerstone tests of the Standard Model~(SM) with unique sensitivity to beyond-the-SM~(BSM) phenomena.
The leading contribution to $a_\ell$ in the SM is the one-loop Schwinger term $\alpha_{_\text{EM}} / 2\pi \simeq 0.00116$~\cite{Schwinger1948,Kusch1948}, where $\alpha_{_\text{EM}}$ is the electromagnetic~(EM) fine-structure constant.
For the electron (muon), $a_e$ ($a_\mu$) is tested to parts per $10^{10}$~\cite{Odom:2006,Hanneke:2010au,Bouchendira:2011,Aoyama:2012wj,Parker191,Morel2020} ($10^{7}$~\cite{Bennett:2006fi,Abi:2021gix,Albahri:2021ixb}) precision.
Measurements of $a_\mu$ report tensions with the SM expectation~\cite{Aoyama:2012wk,Keshavarzi:2018mgv,Davier:2019can,Aoyama:2020ynm,Borsanyi:2020mff}, which may suggest BSM dynamics.
Specific BSM scenarios such as supersymmetry~\cite{Martin:2001st} predict enhancements that scale quadratically with lepton mass $m_\ell$, i.e.\ $\delta a_\ell \propto m_\ell^2$, resulting in a $(m_\tau/m_\mu)^2 \simeq 280$ times larger effect for $\tau$-leptons.
However, the short $\tau$-lepton lifetime precludes precise spin-precession measurements of $a_\tau$ to test the SM prediction of $a_{\tau}^\text{SM} = 0.001\,177\,21~(5)$~\cite{Eidelman:2007sb} and potential BSM contributions.
 
Photon-induced events arise from interactions between the EM fields surrounding the beam particles at colliders.
Observing photon-induced $\tau$-lepton pairs ($\gamma\gamma \to \tau\tau$) predicted to occur at the Large Hadron Collider~(LHC)~\cite{BreitWheeler1934,HeisenbergEuler1936,SchwingerLimit,PhysRevD.7.3485,Klein:2020fmr,Budnev:1974de,Baltz:2007kq,deFavereaudeJeneret:2009db} would open the way to hadron-collider probes of $a_\tau$.
Currently, the most precise single-experiment measurement is $a_\tau = -0.018~(17)$ by the DELPHI Collaboration~\cite{Abdallah:2003xd,Tanabashi:2018oca} using $\gamma\gamma \to \tau\tau$ events at the Large Electron Positron~(LEP) collider.
The OPAL~\cite{Ackerstaff:1998mt} and L3~\cite{Acciarri:1998iv} Collaborations also set constraints using radiative $\tau$-lepton decays~\cite{Gau:1997cn}.
At the LHC, photon-induced dilepton production has only been measured in the dielectron~($ee$) and dimuon~($\mu\mu$) channels, using proton--proton~(\pp)~\cite{STDM-2014-13,STDM-2016-13,STDM-2018-16,CMS-PPS-17-001,CMS-FWD-11-004,CMS-FWD-10-005} and lead--lead~(\PbPb) collisions~\cite{HION-2018-11,HION-2018-19,HION-2016-02,CMS-FSQ-16-012,CMS-HIN-19-014-customRef,ALICE:2013wjo}.
The $\tau\tau$ channel is challenging due to hadronic backgrounds and neutrinos in $\tau$-lepton decays diluting visible final-state kinematics.
This renders triggering and reconstruction more difficult, especially in high-luminosity \pp\ collisions.
Strategies to overcome these experimental obstacles using heavy-ion collisions were proposed in Refs.~\cite{DELAGUILA1991256,Beresford:2019gww,Dyndal:2020yen}.
 
This Letter presents the observation of the $\text{\PbPb} \to \text{Pb}(\gamma\gamma \to \tau\tau)\text{Pb}$
process and measurement of $a_\tau$
using 1.44~nb$^{-1}$ of $\sqn = 5.02$~\TeV\ \PbPb~data recorded by ATLAS in 2018.
The EM fields accompanying the ions coherently create photons that interact to produce $\tau$-lepton pairs.
The cross-section is enhanced by $Z^4$ relative to \pp~collisions, where $Z$ is the atomic number~($Z=82$ for lead).
The ions can remain intact, enabling selection of low-multiplicity events with one muon originating from one of the $\tau$-leptons, while the other $\tau$-lepton decay is reconstructed as either an electron or one or three charged-particle tracks with low transverse momentum.


 
The ATLAS experiment~\cite{PERF-2007-01,ATLAS-TDR-2010-19,PIX-2018-001} is a multipurpose particle detector with cylindrical geometry~\cite{atlas-footnote}, comprising an inner detector~(ID) tracker, EM and hadronic calorimeters, and a muon spectrometer~(MS).
The zero-degree calorimeters~(ZDC)~\cite{ATLAS-TDR-ZDC} are located at $z=\pm 140$~m from the interaction point, and detect neutral particles such as neutrons emitted from interacting nuclei.
A two-level trigger system~\cite{TRIG-2016-01,TRIG-2019-04} was used to select events containing one muon with $\pt > 4$~\GeV, and at most 50~\GeV~(3~\GeV) of transverse energy deposited in the whole (forward $3.2 < |\eta| < 4.9$) calorimeter~\cite{ATL-DAQ-PUB-2019-001-customRef}.
An extensive software suite~\cite{ATL-SOFT-PUB-2021-001} is used in the reconstruction and analysis of real and simulated data, in detector operations, and in the trigger and data acquisition systems.
Standard data-quality requirements are imposed~\cite{DAPR-2018-01}.
The average number of hadronic interactions per bunch crossing was 0.003.
 
Samples of simulated $\gamma\gamma\to \tau\tau$ signal events were produced at leading order in QED using the \textsc{Starlight}~2.0~\cite{Klein:2016yzr} Monte Carlo~(MC) generator, interfaced with
\textsc{Tauola}~\cite{Jadach:1993hs, Davidson:2010rw} for $\tau$-lepton decays.
Final-state radiation (FSR) from the $\tau$-leptons and charged decay products of $\tau$-leptons was simulated using \PYTHIA[8.245]~\cite{Sjostrand:2014zea} and \PHOTOSpp[3.61]~\cite{Davidson:2010ew-customRef}, respectively.
One of the dominant background sources is the $\gamma\gamma\to \mu\mu$ process, and its contribution is estimated with the aid of MC samples generated using \textsc{Starlight};
\PYTHIA[8] was used to model EM FSR from the muons.
The photon-flux distribution in simulated $\gamma\gamma\to \tau\tau$ and $\gamma\gamma\to \mu\mu$ events was reweighted to that of~\textsc{Superchic}~3.05~\cite{Harland-Lang:2018iur}, differentially in dilepton invariant mass and dilepton rapidity.
In the \textsc{Starlight} and \textsc{Superchic} simulations, no restriction on the Coulomb breakup of either nucleus was imposed.
Dijet samples from photon-induced diquark production, $\gamma\gamma\to q\bar{q}$, were generated using \PYTHIA[8].
Nondiffractive photonuclear events ($\gamma\text{A} \to X$) were simulated with \textsc{Starlight} interfaced with \textsc{Dpmjet-III}~\cite{Roesler:2000he}.
All MC samples were passed through a detailed detector simulation based on \GEANT~\cite{Agostinelli:2002hh, SOFT-2010-01}.
 
Charged-particle tracks reconstructed in the ID must satisfy $\pt > 100$~\MeV, $|\eta|<2.5$, transverse impact parameter $\left|d_0\right| < 1.5$~mm, and the ``Loose Primary'' track selection criterion~\cite{ATL-PHYS-PUB-2015-051,PERF-2015-07,PERF-2015-08}.
Electrons must satisfy $\pt > 4$~\GeV, $|\eta|<2.47$ (excluding the calorimeter transition region, i.e.\ $|\eta| \not\in [1.37, 1.52]$), $\left|d_0\right| < 0.5$~mm, and ``Loose''~\cite{EGAM-2018-01} likelihood-based identification criteria.
Muons must satisfy $\pt > 4$~\GeV, $|\eta| < 2.4$, $\left|d_0\right| < 0.3$~mm, and ``LowPt''~\cite{MUON-2018-03} identification criteria.
Small corrections, derived using tag-and-probe methods similar to those in Refs.~\cite{TRIG-2018-01,HION-2016-02,HION-2019-08}, are applied to simulated reconstruction and trigger efficiencies of electrons and muons.
Clusters of topologically connected calorimeter cells called topoclusters~\cite{PERF-2014-07} must satisfy $|\eta| < 4.9$, $\pt>0.1$~\GeV\ ($\pt>1$~\GeV) for $2.5<|\eta|<4.9$ ($|\eta| < 2.5$),
and the cell significance criteria for measured energies outlined in Ref.~\cite{STDM-2011-01}.
Topoclusters from calorimeter regions with an abnormal noise distribution
are removed using a data-driven procedure based on analyzing $\eta$--$\phi$ distributions of topocluster activity for each calorimeter layer.
Reconstructed photons must satisfy $\et > 1.5$~\GeV, $|\eta|<2.37~\not\in [1.37, 1.52]$, and dedicated identification criteria defined in Ref.~\cite{HION-2019-08}.


 
Selected events must contain exactly one muon, which targets a muonic decay of one of the $\tau$-leptons while reducing backgrounds from $\gamma\gamma\to\mu\mu$ and $\gamma\gamma\to q\bar{q}$.
Three signal regions~(SR) then categorize events by the decay signature of the other $\tau$-lepton.
The $\mu e$-SR category additionally requires one electron, and no additional tracks separated from the muon (electron) by $\Delta R_{\mu(e), \text{trk}}>0.1$, which targets fully leptonic decays of both $\tau$-leptons.
The different-flavor ($\mu e$) requirement suppresses same-flavor backgrounds dominated by $\gamma\gamma\to\mu\mu/ee$.
The $\mu$1T-SR ($\mu$3T-SR) category requires exactly one track (three tracks) separated from the muon by $\Delta R_{\mu, \text{trk}}>0.1$, which targets $\tau$-lepton decays to one or three charged hadrons.
The one-track requirement also captures leptonic $\tau$-lepton decays that fail electron or muon reconstruction.
The electric charges of the muon, electron, and tracks must sum to zero.
 
For both $\mu$1T-SR and $\mu$3T-SR, events must contain no additional muons satisfying looser criteria and no electrons to reject $\gamma\gamma\to\mu\mu/ee$ backgrounds.
The looser requirements on muons comprise matched tracks in the ID and MS satisfying $\pt > 2$~\GeV\ and $|\eta| <2.5$.
To suppress hadronic backgrounds such as photonuclear processes, there must be no topoclusters separated from the muon (track or three-track system~\cite{trksystem-def}) by $\Delta R_{\text{clust}, \mu}>0.3$ ($\Delta R_{\text{clust}, \text{trk(s)}}>1.0$); this requirement is referred to as the topocluster veto.
To further reduce photonuclear backgrounds, the acoplanarity between the muon and the track (three-track system) must satisfy $A_{\phi}^{\mu, \text{trk(s)}} \equiv 1 - |\Delta\phi_{\mu, \text{trk(s)}}| / \pi < 0.4~(0.2)$.
The signal has a narrower $A_{\phi}^{\mu, \text{trk(s)}}$ distribution in $\mu$3T-SR than $\mu$1T-SR, motivating the tighter requirement.
 
For $\mu$1T-SR, the \pt\ of the muon--track pair must satisfy $\pt^{\mu,\text{trk}} > 1$~\GeV\ to reject \pt-balanced backgrounds, such as $\gamma\gamma\to\mu\mu$.
To further reduce the $\gamma\gamma\to\mu\mu \gamma$ background, the \pt\ of the muon, track, and photon (topocluster) system must fulfill $\pt^{\mu,\text{trk},\gamma(\text{clust})} > 1$~\GeV\ for events containing a photon (topocluster) within $\Delta R_{\gamma (\text{clust}), \text{trk} }=1$ of the track.
If there are multiple nearby photons (topoclusters), the highest-\pt\ photon (topocluster) is used.
The topoclusters considered here must have $\pt > 2$~\GeV\ and not be track-matched; these criteria avoid track-induced topoclusters from, e.g., charged-pion energy deposits.
Topoclusters within $\Delta R_{\text{clust},\text{trk}}=0.1$ of a track with $\pt > 0.7$~\GeV\ extrapolated to the calorimeter are considered track-matched.
Low-multiplicity events in minimum-bias data are used to correct for the bending due to the magnetic field, such that the $\Delta R_{\text{clust},\text{trk}}$ distribution of topoclusters associated with a track peaks at zero.
Tracks with $\pt < 0.7$~\GeV\ are not considered in the track--cluster matching as they typically do not deposit significant energy in the calorimeter.
 
For $\mu$3T-SR, the three-track system mass must fulfill $m_\text{3trk} < 1.7$~\GeV, assuming each track has the charged-pion mass of $140~\MeV$.
This requirement retains three-prong hadronic $\tau$-lepton decays and suppresses background from exclusive $\rho^0$ mesons ($\gamma\text{A} \to \rho^0 \to \pi\pi$) produced simultaneously with $\gamma\gamma\to\mu\mu$ events.
In this background process, neither muon is correlated with the $\pi\pi$ system and $m_{\pi\pi\mu}$ typically exceeds 1.7~\GeV\ for a muon with \pT of several \GeV.
 
To constrain the $\gamma\gamma\to\mu\mu$ background, a control region (CR) of dimuon events called 2$\mu$-CR is defined.
It requires exactly two muons with invariant mass above 11~\GeV\ to suppress quarkonia ($\Upsilon(nS)\to\mu\mu$) backgrounds and no additional tracks separated from the muons by $\Delta R_{\mu,\text{trk}} > 0.1$.

Events must additionally not have ZDC energies satisfying $E_\text{ZDC} > 1$~\TeV\ on each side, mainly to suppress photonuclear backgrounds where ion dissociation typically occurs.
This class of events with no forward neutrons detected~($0n0n$) corresponds to the absence of Coulomb breakup of either nucleus.
Such breakup typically proceeds through the giant dipole resonance and induces the emission of one or more neutrons~\cite{HION-2016-02}.
This requirement also fully suppresses lepton-pair production in which the initial photon emission results in the dissociation of one or both nuclei~\cite{HION-2016-02}.
The SRs and 2$\mu$-CR are all statistically independent.
Since the extra forward neutron emissions are not simulated, the $\gamma\gamma\to\ell\ell$ MC samples are corrected using data-driven probabilities for the $0n0n$ event topology, which are found to be between 0.4 and 0.7.
These are extracted from 2$\mu$-CR without the $E_\text{ZDC}$ requirement, differentially in dilepton invariant mass and dilepton rapidity.


 
The dominant sources of background after event selection are radiative dimuon ($\gamma\gamma \to \mu\mu \gamma$) and photonuclear events with low central-detector activity.
 
The $\gamma\gamma \to \mu\mu \gamma$ background is estimated with the aid of MC samples.
This process enters $\mu$1T-SR when FSR photons substantially modify the dimuon kinematics so as to mimic the signal kinematics, and enters $\mu 3$T-SR and $\mu e$-SR primarily when photons convert to $e^+e^-$ in detector material.
To improve the modeling of high-\pt~($\pT^\gamma \gtrsim \pT^\mu$) photon emissions for $t$- and $u$-channel $\gamma\gamma \to \mu\mu$ processes, an additional $\gamma\gamma \to \mu\mu\gamma$ MC sample generated using \MGNLO~\cite{Alwall:2014hca} with the photon flux reweighted to \textsc{Superchic} is used instead of \textsc{Starlight}+\PYTHIA[8] if a leading photon has $\pT^{\gamma} > 2$~\GeV.
Comparing the $\gamma\gamma \to \mu\mu(\gamma)$ simulated events with data in $2\mu$-CR shows reasonable data-to-MC agreement in differential distributions, and normalization to within 5\%.
The simulated $\gamma\gamma \to \mu\mu(\gamma)$ event yield in $2\mu$-CR is 15\% lower than data if \textsc{Starlight} photon-flux calculations are used due to known limitations of \textsc{Starlight}~\cite{HION-2016-02,Harland-Lang:2021ysd}.
Before the fit to data, this method estimates 70, 6.5, and 2.8 $\gamma\gamma \to \mu\mu(\gamma)$ events enter $\mu$1T-SR, $\mu 3$T-SR, and $\mu e$-SR, respectively.
 
Diffractive photonuclear backgrounds with low particle activity are estimated using fully data-driven methods.
Dedicated CRs are introduced, called $\mu 2$T-CR ($\mu 4$T-CR), which apply the same selection as $\mu$1T-SR ($\mu 3$T-SR), but require an additional track satisfying $\pT < 0.5$~\GeV.
Furthermore, the $E_\text{ZDC} < 1$~\TeV\ requirement is removed on either side to enrich the sample with events from photonuclear processes.
To suppress the $\gamma\gamma \to \tau\tau$ signal contamination in $\mu 2$T-CR, the two-track system mass must fulfill $m_\text{2trk} > 1$~\GeV; if $m_\text{2trk} < 1$~\GeV, the acoplanarity of the muon and highest-\pt\ track is required to exceed 0.2.
The event yields in CRs are extrapolated to SRs by loosening the veto on topoclusters not matched to the muon or tracks from $n_{\textrm{TC}}^{\textrm{unmatch}} = 0$ to
$n_{\textrm{TC}}^{\textrm{unmatch}} \leq 8$, both in CRs and SRs.
The $\mu$2T-CR ($\mu 4$T-CR) templates for $n_{\textrm{TC}}^{\textrm{unmatch}}$ distributions are normalized to the event yield in $\mu$1T-SR ($\mu 3$T-SR) in the region $4\leq n_{\textrm{TC}}^{\textrm{unmatch}}\leq 8$.
In this region, the signal and dimuon background contributions are found to be negligible, and events exhibit properties that suggest nonexclusive diffractive production, such as a small or no rapidity gap~\cite{rapidity-gap}.
As the additional track in $\mu 2$T-CR and $\mu 4$T-CR is soft ($\pT < 0.5$~\GeV), its possible correlation with topocluster activity is very small.
This method estimates that 13 (2.8) photonuclear events enter $\mu$1T-SR ($\mu 3$T-SR); photonuclear events are expected to be negligible in $\mu e$-SR.
 
Other sources of background are predicted to be negligible in the SRs.
Nondiffractive photonuclear interactions are estimated using the \textsc{Starlight+Dpmjet-III} sample and are found to be negligible in all three SRs.
The \PYTHIA[8] simulation of $\gamma\gamma \to q\bar{q}$ estimates dijet backgrounds contribute less than 0.3 events in both $\mu $1T-SR and $\mu 3$T-SR.
Similarly, the contribution from resolved $\gamma\gamma$ interactions, as estimated with \PYTHIA[8]~\cite{Helenius:2017aqz}, is found to be negligible.
Exclusive $\rho^0$-meson production with simultaneous $\gamma\gamma \to \mu\mu$ production in $\mu 3$T-SR is studied using a data-driven method.
Template distributions for this process are built from events with two muons and two additional charged-particle tracks.
The acoplanarities of both the muon pair and the track pair must be below 0.05.
All events in these templates are found to have $m_\text{3trk} > 1.7$~\GeV, so this background is expected to be negligible in $\mu 3$T-SR.


 
Systematic uncertainties affecting the measurement arise from the reconstruction of leptons, photons, charged-particle tracks and topoclusters, the signal and background modeling, and integrated luminosity.
 
Uncertainties in the muon momentum scale and resolution follow those in Ref.~\cite{PERF-2015-10}.
The analysis includes uncertainties in the data-to-MC correction factors applied to simulated samples for the muon trigger and reconstruction efficiencies.
Uncertainties in the reconstruction, identification, and energy calibration of electrons and photons are evaluated in accord with Ref.~\cite{HION-2019-08}.
The uncertainty in the inclusive track reconstruction efficiency is dominated by the uncertainty in the amount of ID material~\cite{PERF-2015-07}.
This uncertainty is applied in the simulation by randomly removing tracks with a \pt- and $\eta$-dependent probability corresponding to the material uncertainty.
Uncertainties in the topocluster reconstruction efficiency and energy calibration are estimated using $\gamma\gamma\to e e$ events where one of the electrons emits a hard bremsstrahlung photon due to its interaction with detector material~\cite{HION-2019-08}.
 
Uncertainties in the photonuclear background evaluation are estimated by repeating the procedure with alternative requirements for CRs.
These resemble the $\mu 1$T-SR ($\mu 3$T-SR) selection except that the track (three-track system) has the same electric charge as the muon candidate.
The difference between the photonuclear background contribution evaluated with alternative and nominal CRs defines the uncertainties, affecting both the normalization and differential distributions.
Uncertainties in modeling the photon flux are estimated by using the \textsc{Starlight} MC samples without reweighting to \textsc{Superchic}.
This affects the normalization and differential distributions of the signal and $\gamma\gamma\to \mu\mu$ background.
Uncertainties in modeling $\tau$-lepton decays are estimated using \PYTHIA[8]~\cite{Ilten:2015mja} as an alternative MC simulation to \textsc{Tauola}.
The effect of $\tau$-lepton spin correlations in \textsc{Tauola} is implemented using helicity amplitudes from the $\gamma^*\to \tau \tau$ process.
This modeling is therefore cross-checked by comparing signal events simulated using two versions of \PYTHIA[8]: v8.245 uses helicity amplitudes from the $\gamma^*\to \tau \tau$ process, whereas v8.305 uses the $\gamma\gamma\to \tau \tau$ elementary process.
The difference between the two implementations is found to be negligible, and no further systematic uncertainty is assigned.
 
The uncertainty in the integrated luminosity is 1.9\%, obtained with the LUCID-2 detector~\cite{LUCID2} using methods similar to Ref.~\cite{ATLAS-CONF-2019-021} for the primary luminosity measurements.


After applying the event selection, a total of 532, 85, and 39 data events are observed, compared with $84 \pm 19$, $9 \pm 3$, and $2.8 \pm 0.7$ expected background events in $\mu$1T-SR, $\mu$3T-SR, and $\mu e$-SR, respectively.
The background-only hypothesis is rejected with significance exceeding $5\sigma$, establishing the observation of the $\gamma\gamma \to \tau\tau$ process at ATLAS.
The signal significance is highest in $\mu$1T-SR, while $\mu e$-SR has the largest signal-to-background ratio.
The pre-fit signal-plus-background hypothesis predicts $543 \pm 111$, $93 \pm 20$, and $35 \pm 8$ events in $\mu$1T-SR, $\mu$3T-SR, and $\mu e$-SR, respectively, which is compatible with the observed data.
The signal strength $\mu_{\tau\tau}$, defined as the ratio of the observed signal yield to the SM expectation assuming the SM value for $a_{\tau}$, is measured using a profile-likelihood fit~\cite{Cowan:2010js,Cranmer:1456844} to be $\mu_{\tau\tau} = 1.03^{+0.06}_{-0.05}~\textrm{(tot)} = 1.03^{+0.05}_{-0.05}~\textrm{(stat)}~^{+0.03}_{-0.03}~\textrm{(syst)}$.
The fit uses the $\pT^\mu$ distribution in the three SRs and $2\mu$-CR with $\mu_{\tau\tau}$ being the only parameter of interest.

Approximately 80 nuisance parameters representing the systematic uncertainties are included in the fit.
Many systematic uncertainties are correlated between the SRs and $2\mu$-CR, so their impact on the measurement precision is minimized since they are constrained by $2\mu$-CR.
The dominant pre-fit contribution is the photon-flux uncertainty, which mainly affects the signal yield (by approximately 20\%), with a significantly smaller impact on the signal shape found upon decorrelation from the normalization component.
After the fit, the photon-flux uncertainty becomes subdominant and luminosity uncertainty becomes negligible relative to other sources.
The leading contributions to the total systematic uncertainty are the estimation of the muon trigger efficiency, $\tau$-lepton decay modeling, and track reconstruction efficiency.

To measure $a_{\tau}$, an alternative fit is performed where $a_{\tau}$ is the only free parameter using the $\pT^\mu$ distribution in the three SRs and $2\mu$-CR;
$\pT^\mu$ is chosen because of its high sensitivity to $a_{\tau}$~\cite{Dyndal:2020yen}.
Simulated signal samples with various $a_\tau$ values are employed.
In the nominal sample, $a_\tau$ is set to its SM value.
Signal templates for alternative $a_\tau$ hypotheses are obtained by reweighting the nominal sample in three dimensions, differentially in $\tau\tau$ invariant mass, $\tau\tau$ rapidity, and rapidity difference between the two $\tau$-leptons, according to calculations from Ref.~\cite{Dyndal:2020yen}.
These calculations parameterize the $\tau\tau\gamma$ coupling by $F_1(q^2)\gamma^{\mu} + F_2(q^2)\frac{i}{2m_{\tau}}\sigma^{\mu\nu}q_{\nu}$, where $q_{\nu}$ is the photon four-momentum, $\sigma^{\mu\nu} = \text{i}[\gamma^\mu, \gamma^\nu]/2$ the spin tensor, and the form factors satisfy $F_1(q^2\to 0)=1$ and $F_2(q^2\to 0)=a_{\tau}$.
A similar parameterization was used in previous LEP measurements~\cite{Ackerstaff:1998mt,Acciarri:1998iv,Abdallah:2003xd}, which exploits the near-zero virtuality of initial-state photons.
A total of 14 templates for different $a_{\tau}$ values are created to model the dependence of the $\pT^\mu$ distribution on $a_\tau$ in the three SRs.
 
\begin{figure}
\begin{center}
\includegraphics[width=0.45\textwidth]{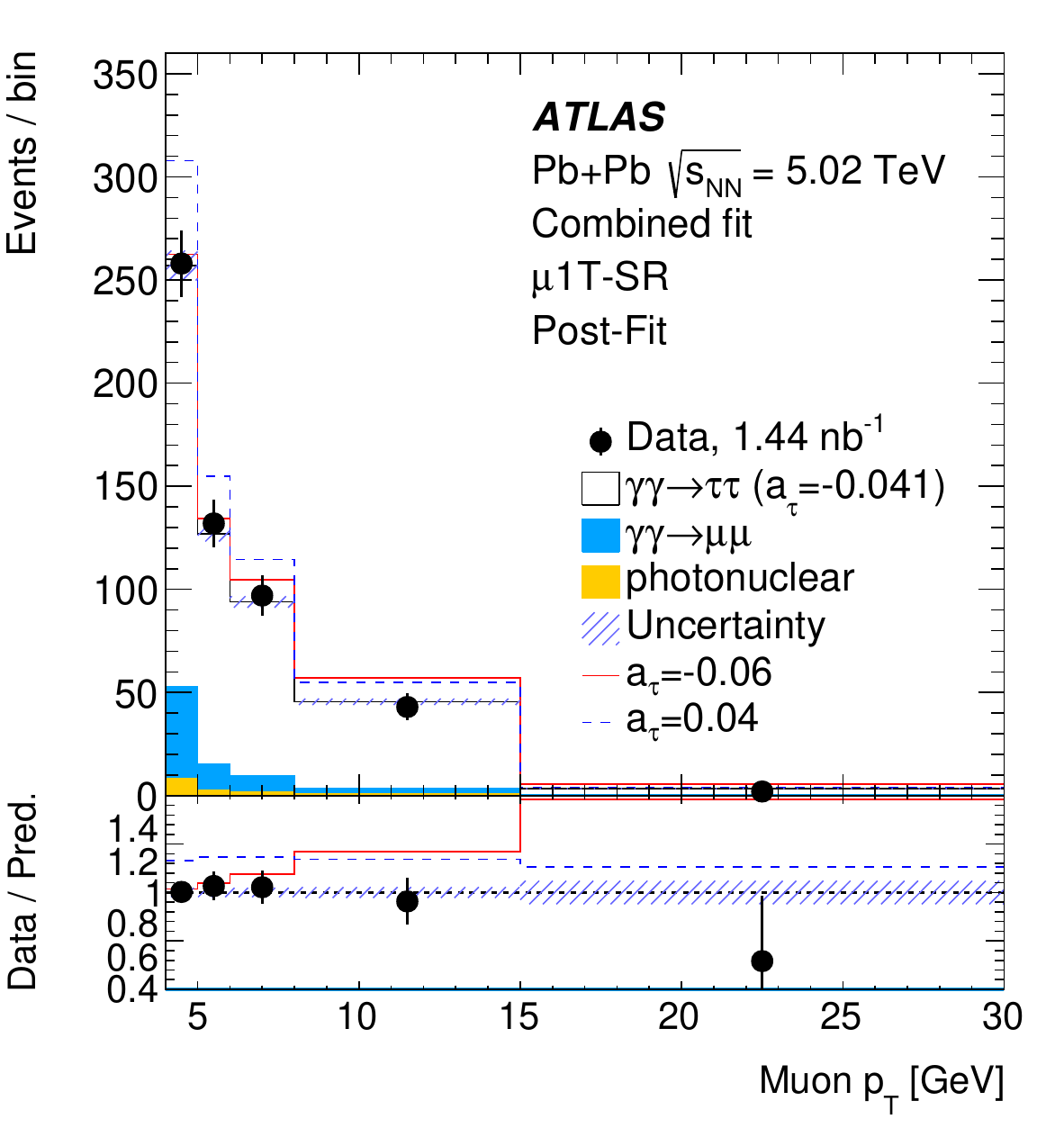}
\includegraphics[width=0.45\textwidth]{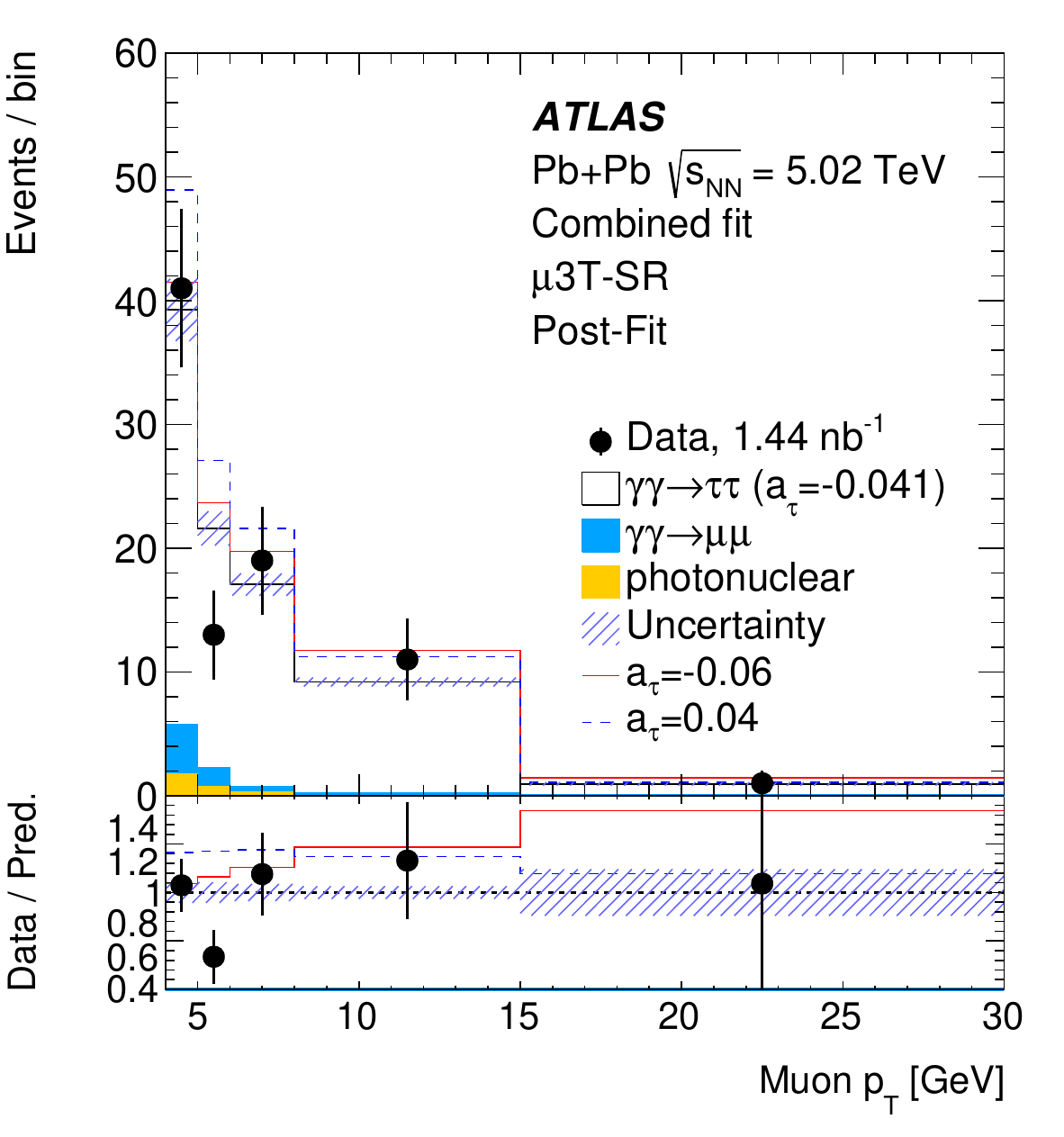}\\
\includegraphics[width=0.45\textwidth]{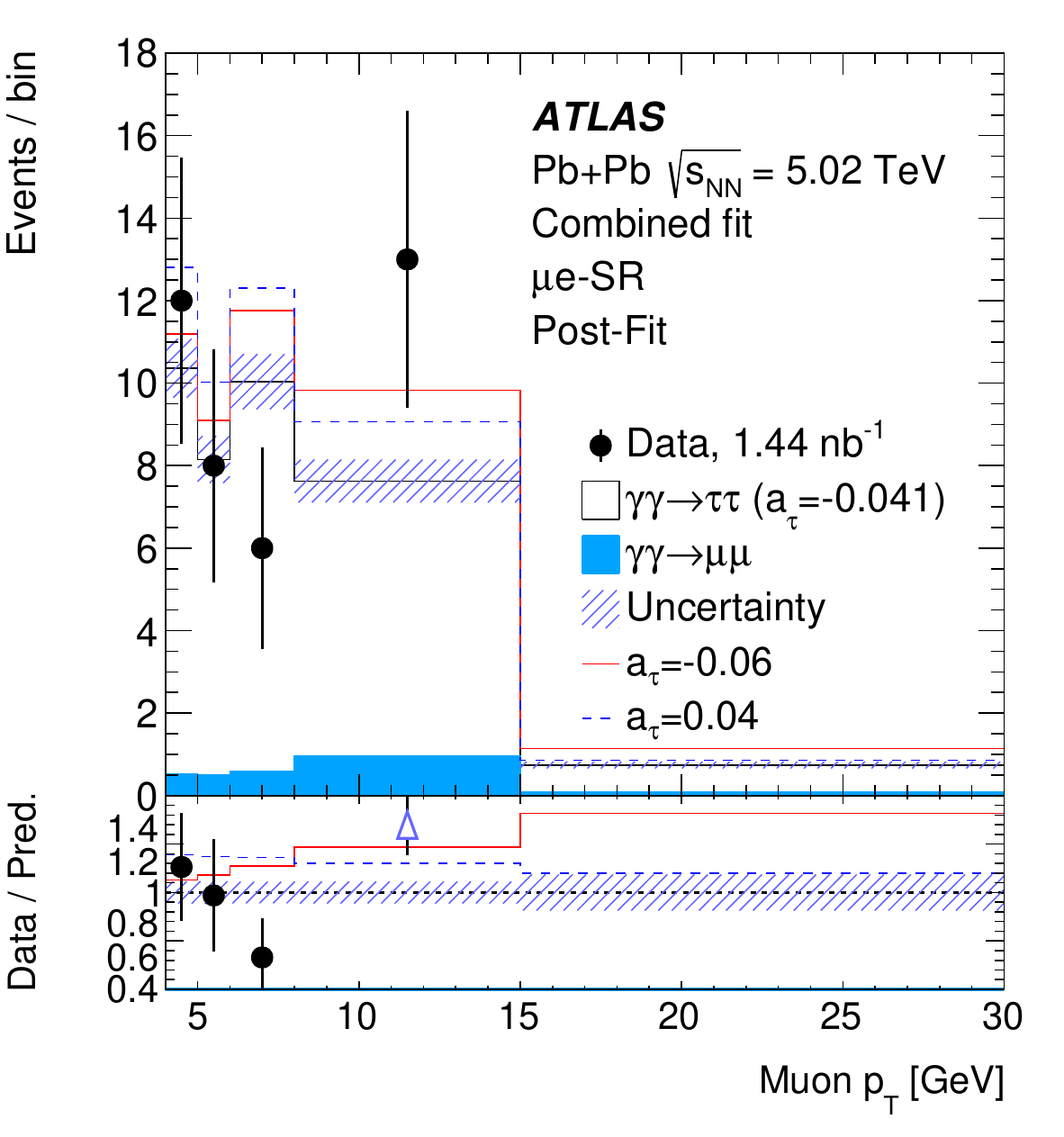}
\includegraphics[width=0.45\textwidth]{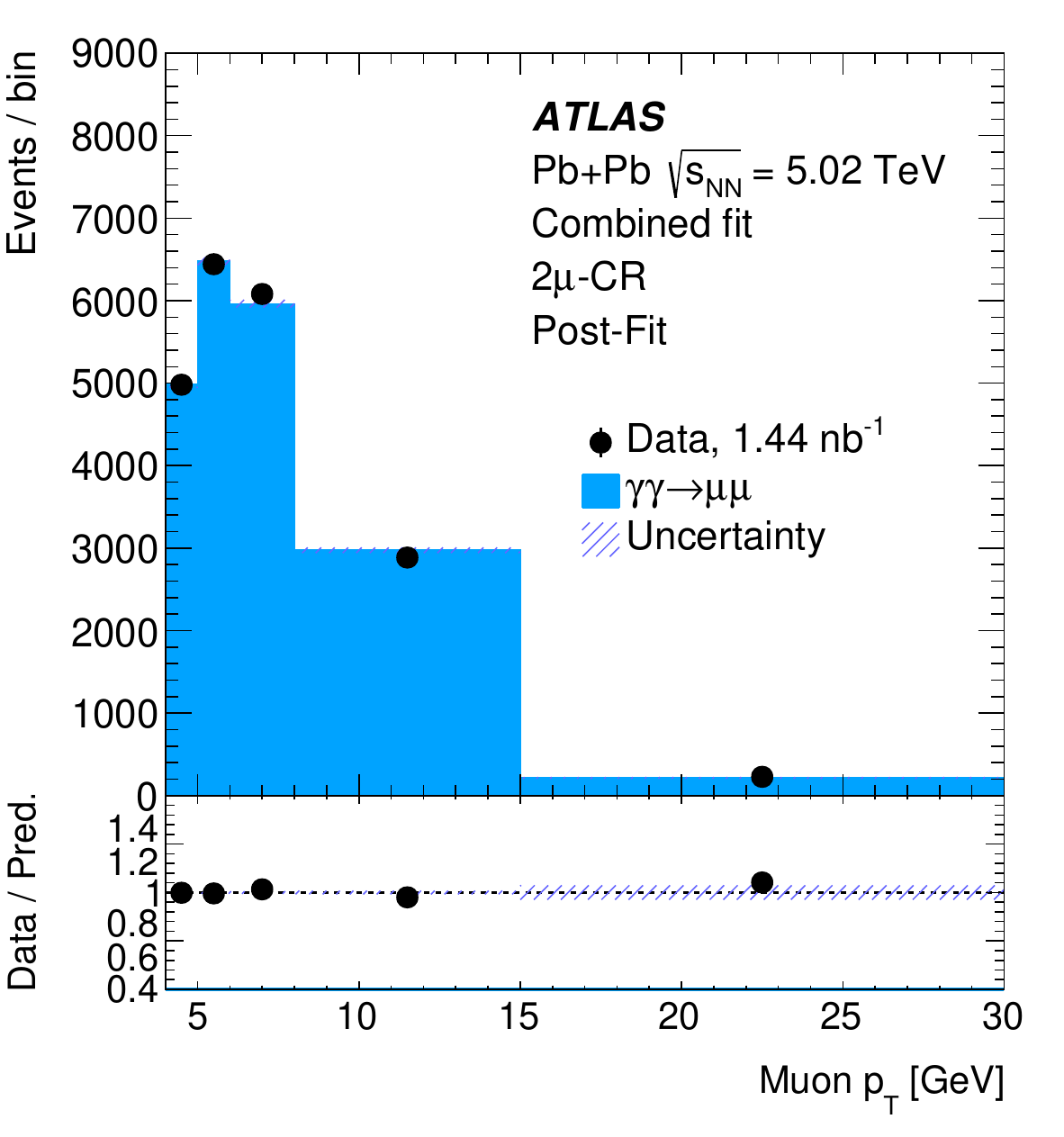}\\
\end{center}
\caption{Muon transverse momentum distributions in the (top-left) $\mu$1T-SR, (top-right) $\mu$3T-SR, (bottom-left) $\mu$e-SR, and (bottom-right) $2\mu$-CR categories.
Black markers denote data and stacked histograms indicate the different components contributing to the regions.
Post-fit distributions are shown with the signal contribution corresponding to the best-fit $a_{\tau}$ value ($a_{\tau}=-0.041$).
For comparison, signal contributions with alternative $a_{\tau}$ values are shown as solid red ($a_{\tau}=-0.06$) or dashed blue ($a_{\tau}=0.04$) lines.
The bottom panel shows the ratio of the data to post-fit predictions.
Vertical bars denote uncertainties from the finite number of data events.
Hatched bands represent $\pm 1\sigma$ systematic uncertainties of the prediction with the constraints from the fit applied.}
\label{fig:postFit}
\end{figure}

Figure~\ref{fig:postFit} shows the $\pT^\mu$ distributions of the four analysis regions for the data and post-fit expectation.
The fit describes the data well.
 
\begin{figure}
\begin{center}
\includegraphics[width=0.7\textwidth]{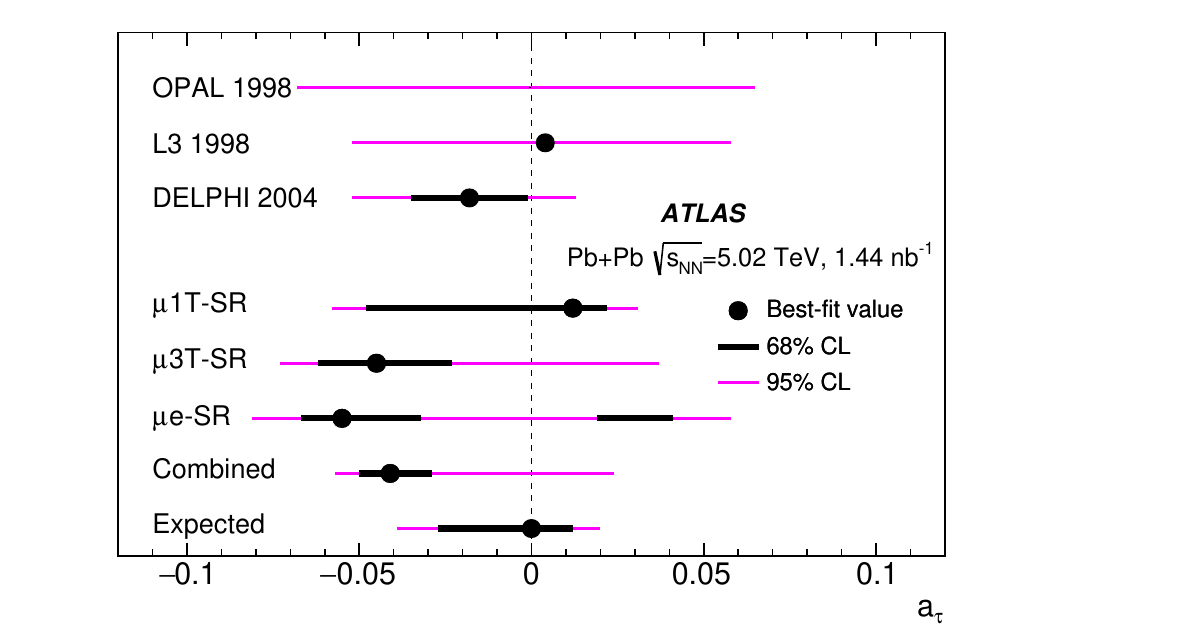}
\end{center}
\caption{Measurements of $a_{\tau}$ from fits to individual signal regions (including the dimuon control region), and from the combined fit.
These are compared with existing measurements from the OPAL~\cite{Ackerstaff:1998mt}, L3~\cite{Acciarri:1998iv} and DELPHI~\cite{Abdallah:2003xd} experiments at LEP.
A point denotes the best-fit $a_{\tau}$ value for each measurement if available, while thick black (thin magenta) lines show 68\% CL (95\% CL) intervals.
The expected interval from the ATLAS combined fit is also shown.}
\label{fig:limits}
\end{figure}

The best-fit value of $a_{\tau}$ is $a_{\tau} = -0.041$, with the corresponding 68\% CL and 95\% CL intervals being $(-0.050,~-0.029)$ and $(-0.057,~0.024)$, respectively.
The higher-than-expected observed yields lead to the highly asymmetric 95\% CL interval.
This arises from the nearly quadratic signal cross-section dependence on $a_{\tau}$, caused by the interference of the SM and BSM amplitudes~\cite{Ackerstaff:1998mt,Acciarri:1998iv,Dyndal:2020yen}.
The expected 95\% CL interval is $-0.039 < a_\tau < 0.020$.
The impact of systematic uncertainties on the final results is small relative to statistical uncertainties.
Figure~\ref{fig:limits} shows the $a_{\tau}$ measurement alongside previous results obtained at LEP.
The precision of this measurement is similar to the most precise single-experiment measurement by the DELPHI Collaboration.
 
\FloatBarrier

In summary, $\tau$-lepton pair production in ultraperipheral heavy-ion collisions, $\text{\PbPb} \to \text{Pb}(\gamma\gamma \to \tau\tau)\text{Pb}$, is observed by ATLAS with a significance exceeding $5\sigma$ in 1.44~nb$^{-1}$ of $\sqn = 5.02$~\TeV\ data at the LHC.
The observed event yield is compatible with that expected from the SM prediction within uncertainties.
The events are used to set constraints on the $\tau$-lepton anomalous magnetic moment, corresponding to $-0.057<a_\tau <0.024$ at 95\% CL.
The measurement precision is limited by statistical uncertainties.
This result introduces the use of hadron-collider data to test electromagnetic properties of the $\tau$-lepton, and the results are competitive with existing lepton-collider constraints.


\section*{Acknowledgments}


We thank CERN for the very successful operation of the LHC, as well as the
support staff from our institutions without whom ATLAS could not be
operated efficiently.
 
We acknowledge the support of
ANPCyT, Argentina;
YerPhI, Armenia;
ARC, Australia;
BMWFW and FWF, Austria;
ANAS, Azerbaijan;
CNPq and FAPESP, Brazil;
NSERC, NRC and CFI, Canada;
CERN;
ANID, Chile;
CAS, MOST and NSFC, China;
Minciencias, Colombia;
MEYS CR, Czech Republic;
DNRF and DNSRC, Denmark;
IN2P3-CNRS and CEA-DRF/IRFU, France;
SRNSFG, Georgia;
BMBF, HGF and MPG, Germany;
GSRI, Greece;
RGC and Hong Kong SAR, China;
ISF and Benoziyo Center, Israel;
INFN, Italy;
MEXT and JSPS, Japan;
CNRST, Morocco;
NWO, Netherlands;
RCN, Norway;
MEiN, Poland;
FCT, Portugal;
MNE/IFA, Romania;
MESTD, Serbia;
MSSR, Slovakia;
ARRS and MIZ\v{S}, Slovenia;
DSI/NRF, South Africa;
MICINN, Spain;
SRC and Wallenberg Foundation, Sweden;
SERI, SNSF and Cantons of Bern and Geneva, Switzerland;
MOST, Taiwan;
TENMAK, T\"urkiye;
STFC, United Kingdom;
DOE and NSF, United States of America.
In addition, individual groups and members have received support from
BCKDF, CANARIE, Compute Canada and CRC, Canada;
PRIMUS 21/SCI/017 and UNCE SCI/013, Czech Republic;
COST, ERC, ERDF, Horizon 2020 and Marie Sk{\l}odowska-Curie Actions, European Union;
Investissements d'Avenir Labex, Investissements d'Avenir Idex and ANR, France;
DFG and AvH Foundation, Germany;
Herakleitos, Thales and Aristeia programmes co-financed by EU-ESF and the Greek NSRF, Greece;
BSF-NSF and MINERVA, Israel;
Norwegian Financial Mechanism 2014-2021, Norway;
NCN and NAWA, Poland;
La Caixa Banking Foundation, CERCA Programme Generalitat de Catalunya and PROMETEO and GenT Programmes Generalitat Valenciana, Spain;
G\"{o}ran Gustafssons Stiftelse, Sweden;
The Royal Society and Leverhulme Trust, United Kingdom.
 
The crucial computing support from all WLCG partners is acknowledged gratefully, in particular from CERN, the ATLAS Tier-1 facilities at TRIUMF (Canada), NDGF (Denmark, Norway, Sweden), CC-IN2P3 (France), KIT/GridKA (Germany), INFN-CNAF (Italy), NL-T1 (Netherlands), PIC (Spain), ASGC (Taiwan), RAL (UK) and BNL (USA), the Tier-2 facilities worldwide and large non-WLCG resource providers. Major contributors of computing resources are listed in Ref.~\cite{ATL-SOFT-PUB-2023-001}.


\printbibliography
 
\clearpage
 
\begin{flushleft}
\hypersetup{urlcolor=black}
{\Large The ATLAS Collaboration}

\bigskip

\AtlasOrcid[0000-0002-6665-4934]{G.~Aad}$^\textrm{\scriptsize 101}$,
\AtlasOrcid[0000-0002-5888-2734]{B.~Abbott}$^\textrm{\scriptsize 119}$,
\AtlasOrcid[0000-0002-7248-3203]{D.C.~Abbott}$^\textrm{\scriptsize 102}$,
\AtlasOrcid[0000-0002-1002-1652]{K.~Abeling}$^\textrm{\scriptsize 55}$,
\AtlasOrcid[0000-0002-8496-9294]{S.H.~Abidi}$^\textrm{\scriptsize 29}$,
\AtlasOrcid[0000-0002-9987-2292]{A.~Aboulhorma}$^\textrm{\scriptsize 35e}$,
\AtlasOrcid[0000-0001-5329-6640]{H.~Abramowicz}$^\textrm{\scriptsize 150}$,
\AtlasOrcid[0000-0002-1599-2896]{H.~Abreu}$^\textrm{\scriptsize 149}$,
\AtlasOrcid[0000-0003-0403-3697]{Y.~Abulaiti}$^\textrm{\scriptsize 116}$,
\AtlasOrcid[0000-0003-0762-7204]{A.C.~Abusleme~Hoffman}$^\textrm{\scriptsize 136a}$
% ...
\bigskip
\\

% ...
$^{29}$Physics Department, Brookhaven National Laboratory, Upton NY; United States of America.\\
% ...
$^{35}$$^{(a)}$Facult\'e des Sciences Ain Chock, R\'eseau Universitaire de Physique des Hautes Energies - Universit\'e Hassan II, Casablanca;$^{(b)}$Facult\'{e} des Sciences, Universit\'{e} Ibn-Tofail, K\'{e}nitra;$^{(c)}$Facult\'e des Sciences Semlalia, Universit\'e Cadi Ayyad, LPHEA-Marrakech;$^{(d)}$LPMR, Facult\'e des Sciences, Universit\'e Mohamed Premier, Oujda;$^{(e)}$Facult\'e des sciences, Universit\'e Mohammed V, Rabat;$^{(f)}$Institute of Applied Physics, Mohammed VI Polytechnic University, Ben Guerir; Morocco.\\
% ...
$^{55}$II. Physikalisches Institut, Georg-August-Universit\"{a}t G\"ottingen, G\"ottingen; Germany.\\
% ...
$^{101}$CPPM, Aix-Marseille Universit\'e, CNRS/IN2P3, Marseille; France.\\
$^{102}$Department of Physics, University of Massachusetts, Amherst MA; United States of America.\\
% ...
$^{116}$Department of Physics, New York University, New York NY; United States of America.\\
% ...
$^{119}$Homer L. Dodge Department of Physics and Astronomy, University of Oklahoma, Norman OK; United States of America.\\
% ...
$^{136}$$^{(a)}$Departamento de F\'isica, Pontificia Universidad Cat\'olica de Chile, Santiago;$^{(b)}$Millennium Institute for Subatomic physics at high energy frontier (SAPHIR), Santiago;$^{(c)}$Instituto de Investigaci\'on Multidisciplinario en Ciencia y Tecnolog\'ia, y Departamento de F\'isica, Universidad de La Serena;$^{(d)}$Universidad Andres Bello, Department of Physics, Santiago;$^{(e)}$Instituto de Alta Investigaci\'on, Universidad de Tarapac\'a, Arica;$^{(f)}$Departamento de F\'isica, Universidad T\'ecnica Federico Santa Mar\'ia, Valpara\'iso; Chile.\\
% ...
$^{149}$Department of Physics, Technion, Israel Institute of Technology, Haifa; Israel.\\
$^{150}$Raymond and Beverly Sackler School of Physics and Astronomy, Tel Aviv University, Tel Aviv; Israel.\\
% ...
$^{a}$ Also Affiliated with an institute covered by a cooperation agreement with CERN.\\
$^{b}$ Also at An-Najah National University, Nablus; Palestine.\\
$^{c}$ Also at Borough of Manhattan Community College, City University of New York, New York NY; United States of America.\\
$^{d}$ Also at Bruno Kessler Foundation, Trento; Italy.\\
$^{e}$ Also at Center for High Energy Physics, Peking University; China.\\
$^{f}$ Also at Centro Studi e Ricerche Enrico Fermi; Italy.\\
% ...

\end{flushleft}


\end{document}